\documentclass[journal]{IEEEtran}

\usepackage{amssymb}
\usepackage{amsmath}
\usepackage{blindtext, graphicx}
\usepackage{tabularx}
\usepackage{multicol}
 \usepackage{relsize}
 \usepackage{float}
 \usepackage{stfloats}
\usepackage{cuted}
\usepackage{cite}
\usepackage{url}
\usepackage{multirow}
\usepackage{hyperref}

\usepackage{xcolor}
\usepackage{soul}
\usepackage{subcaption}

\begin{document}
\pdfoutput=1
%
\title{ Emulating UAV Motion by Utilizing Robotic Arm for  mmWave Wireless Channel Characterization}
%

\author{Amit~Kachroo,~\IEEEmembership{Student~Member,~IEEE,}
        Collin~A.~Thornton, 
        Md~Arifur~Rahman~Sarker,
         Wooyeol~Choi,~\IEEEmembership{Senior~Member,~IEEE,}
         He~Bai,~\IEEEmembership{Member,~IEEE,}
         Ickhyun~Song,~\IEEEmembership{Member,~IEEE,}
         John~O'Hara,~\IEEEmembership{Senior~Member,~IEEE,} and
         Sabit~Ekin$^*$,~\IEEEmembership{Member,~IEEE}
\thanks{ A.~Kachroo, C.~Thornton, Md.~Sarker, W.~Choi, I.~Song, J.~O'Hara and S.~Ekin are with the School of Electrical and Computer Engineering, Oklahoma State University, OK, USA (e-mail: \{amit.kachroo, collin.thornton,  msarker, wchoi, isong, oharaj, sabit.ekin\}@okstate.edu), *Corresponding author: Sabit Ekin.}
\thanks{H.~Bai is with the Department of Mechanical and Aerospace Engineering, Oklahoma State University, OK, USA (email: he.bai@okstate.edu).} 
}

	\markboth{Accepted to IEEE Transactions on Antennas and Propagation, 2021}{Shell \MakeLowercase{\textit{et al.}}: Bare Demo of IEEEtran.cls for Journals}

\maketitle


\begin{abstract}
In this paper, millimeter wave (mmWave) wireless channel characteristics (Doppler spread and path loss modeling) for Unmanned Aerial Vehicles (UAVs) assisted  communication is analyzed and studied by emulating the real UAV motion using a robotic arm. The  motion considers the actual turbulence caused by the wind gusts to the UAV in the atmosphere, which is statistically modeled by the widely used Dryden wind model. The  frequency under consideration is 28 GHz in an anechoic chamber setting. A total of 11 distance points from 3.5 feet to 23.5 feet in increments of 2 feet were considered in this experiment. At each distance point, 3 samples of data were collected for better inference purposes. In this emulated environment, it was found out that the average Doppler spread at these different distances was around -20 Hz and +20 Hz at the noise floor of -60 dB. On the other hand, the path loss exponent was found to be 1.843.  This study presents and lays out  a novel framework of emulating UAV motion for mmWave communication systems, which will  pave the way out for future  design and implementation of next generation UAV-assisted wireless communication systems.
\end{abstract}

\begin{IEEEkeywords}
mmWave, UAV, Doppler, Channel emulation, Path loss, Dryden wind model, Path loss 
\end{IEEEkeywords}

%
\IEEEpeerreviewmaketitle

\section{Introduction}

Rapid proliferation of smart devices below 6 GHz has lead to a massive amount of data traffic, thus creating a tremendous burden on the limited frequency spectrum available to the public. To overcome this spectrum crunch, technologies such as cognitive radios\cite{hong2014cognitive}, multiple-input and multiple-output (MIMO)\cite{liang2014low}, and non-orthogonal multiple access (NOMA)\cite{ding2017surveyNoma} etc. were  proposed. However, the demand still continues to outpace the spectrum availability. In 2015, the Federal Communications Commission (FCC) released millimeter wave (mmWave) frequencies for licensed and unlicensed use. The newly licensed frequencies are 28 GHz, 37 GHz, and 39 GHz, while the unlicensed bands are  64$\sim$71 GHz\cite{5Gspectrum}. Access to these mmWave  frequencies allows multi gigabit wireless communication which enables fifth generation (5G) and beyond fifth generation (B5G) communications\cite{o2019perspective}. In addition to  the high data rates, small antenna size, and circuits at millimeter wavelength  provides reliable, highly directional and secure communication links against any eavesdropping and jamming. 

On the other hand, unmanned aerial vehicles (UAVs), also referred to as drones, have seen a lot of interest from academia, industry, and from the general public at large over the past decade. The main reasons behind this large scale popularity  are the ease of operability (remote or autonomous), easy deployment, higher maneuverability,  and lower operating and maintenance costs of UAVs. UAVs are now extensively used in smart farming, disaster responses, military, smart logistics, and recreation (filming)\cite{zhang2012application,zhang2019survey,erdelj2017help,naqvi2018drone,hayat2016survey}. Advancements in UAV technology are also fuelling the interest of its application in wireless communication technology.  UAVs are capable of  providing a highly reliable and cost effective mode of technology\cite{zhang2019survey,mozaffari2019tutorial,zeng2016wireless,sekander2018multi,vinogradov2019tutorial}, and can also be easily deployed as a flying base station (BS) to provide ubiquitous wireless communication access. They also provide an alternative support for 5G and B5G cellular mobile communication. Furthermore, UAVs can be used as mobile relays to provide wireless connectivity among partitioned  user equipment (UE) that lack any direct line of sight (LOS) communication between the BS and UE. Apart from using as a flying BS and/or relaying node, UAVs have also found application in areas such as aerial data collectors, aerial caching, and aerial power source etc. Multiple UAVs can also coordinate, and self organise to form different network architectures, such as flying adhoc networks (FANETs), internet of drones etc. All these applications of UAV assisted wireless communication have been studied assuming Wi-Fi, or at fourth generation (4G) cellular communication frequencies. Lately, mmWave communication with UAVs have been a topic of great interest\cite{khawaja2019survey,wang2018millimeter,zhang2019survey}. 


Recent studies\cite{wang2018millimeter,rappaport2013millimeter,kachroo2019unmanned,qureshi2019tradeoffs,khawaja2019survey,zhang2019survey} suggest that the successful deployment of UAV-assisted mmWave wireless communication hinges on accurate and realistic propagation channel modeling. While considerable research on terrestrial propagation channels has been conducted for the last few decades, propagation channel modeling for UAVs has not been extensively studied\cite{khawaja2019survey}. Existing channel modeling studies mainly use 1) analytical modeling, e.g., two-ray model, or 2) ray-tracing simulations, or 3) empirical modeling using channel sounding methods. The first two methods are deterministic, cost-effective and require less effort, but they are fundamentally sub-optimal approaches given the complexity and dynamicity of the wireless channels. On the other hand, empirical modeling that uses the channel sounding method could provide more realistic channel models but requires a large amount of statistical data that needs to be collected from  multiple channel observations, and numerous measurement campaigns by using advanced channel sounding equipment. Moreover, microwave measurements to validate such models are inadequate because of fundamental differences between mmWave and microwave channels (e.g., propagation loss, directivity, sensitivity to blockage). The UAV operating environment also introduces unique atmospheric and terrain challenges\cite{khawaja2019survey}. Due to UAV restrictions (e.g., pointing, payload, power, equipment-cost constraints) and the requirement of advanced channel-sounding equipment, to our knowledge no studies have been reported that conduct empirical modeling for UAV-assisted mmWave channels. Available studies mainly use ray-tracing simulations or analytical modeling\cite{qureshi2019tradeoffs,khawaja2019survey}, e.g., for frequencies of 28 GHz and 60 GHz. Also, due to differences in channel scattering environment and operating frequencies, the propagation channel
models used for higher altitude aeronautical communications generally cannot be utilized directly for low altitude (small) UAV-assisted mmWave communications\cite{khawaja2019survey}. Distinct structural and flight characteristics for
low-altitude UAVs could be expected such as different airframe shadowing features, and potentially sharper
pitch, roll, and yaw rates of change during flight. Empirical data is therefore required to determine accurate analytic and stochastic models of mmWave wireless channels.

Utilization of a robotic arm to emulate UAV motion can 1) help overcome these challenges, 2) efficiently emulate the motion of UAV in different environments/scenarios, and 3) enable rapid collection of channel measurements by using channel sounding method to create a database for UAV-assisted mmWave channel models. 
In this work, a novel way to incorporate UAV motion in studying UAV assisted mmWave communication is conducted by emulating the UAV motion by a robotic arm in an anechoic chamber. This method, as compared to other studies which mostly depend on the software simulation, captures the real UAV dynamics in the UAV motion\footnote{Rotor motion is not considered in this scenario, which can be a part of future work.}. This study, therefore, is an important first step in realizing a real mmWave communication system for UAVs in the near future.
Designing a channel emulator by using vector network analyzer (VNA) based channel sounder with real UAV motion emulated by a robotic arm to develop a first-of-its-kind, experimental, wireless channel emulator for UAV-assisted mmWave communications can produce accurate and realistic propagation channel models for a wide range of environments and scenarios. Using the robotic arm to produce atmospheric turbulence effects on UAV position and stability, we can produce the first empirical mmWave channel models that account for path loss/shadowing, Doppler spread due to UAV motion.

Before moving on to the measurement set up in the next section, it is very important to understand the different challenges associated with the mmWave communication itself.  mmWave  communication  suffers extensively from  the propagation attenuation, shadowing effect (blocking), beam misalignment,  and Doppler shift\cite{rappaport2013millimeter, hemadeh2017millimeter,siles2015atmospheric,hassanieh2018fast} because of small wavelength in the order of millimeter. Doppler spread effect is the most critical one, when there is a motion attributed to the movement of transmitter (Tx) or receiver (Rx) or  both. In addition to that, with  the wind gusts in the atmosphere for UAVs, Doppler effects are more aggravated. It is also well known that in a  given mobile  and  multipath environment, each multipath component (MPC) will experience a different Doppler shift according to the motion. This leads to the  spectral broadening at the receiver that causes erroneous signal reception or communication failure. Therefore, modeling physical UAV motion as close to a real life UAV motion  is very crucial in understanding the design constraints, and performance of a  mmWave based UAV communication system.  Techniques to analyze and  combat the Doppler effects  have been studied earlier but as mentioned previously, almost all of them  are  based on simulations that ignore the actual UAV motion dynamics and the wind gust conditions in the atmosphere\cite{khawaja2019survey,zhang2019survey,sekander2018multi}. Therefore, in this paper, the focus is to analyze and study the Doppler and channel power characteristics  by emulating the real UAV motion under  wind gusts. This close to actual UAV motion is emulated by using a robotic arm in an anechoic chamber environment. 

This study will: 1) empirically characterize the time-variant radio propagation channel of a UAV-assisted communication in mmWave range, 2) underpin novel control and aerospace designs for next-generation UAVs and UAV-assisted communications, and 3) provide the stepping stone for future channel emulators of high-complexity to analyze next-generation communication systems. In summary, the main contributions of this paper are as follows,
\begin{itemize}
    \item Doppler spread analysis of mmWave communication system by emulating UAV motion under the Dryden wind conditions in an  anechoic chamber.
    \item De-embedding of phase variations caused by the cable movement from collected data in this VNA based measurement.
    \item Determination of path loss exponent value under anechoic chamber settings\footnote{It is important to note that these results cannot be generalized as such for any indoor or outdoor environments, and will require extensive data measurement campaigns in that regards.}.
\end{itemize}

This paper is organized as follows, Section \ref{meas_setup} discusses the measurement setup. In Section \ref{UAV_motion}, details of the UAV motion emulation is discussed. Analysis of Doppler spread and corresponding signal processing are presented in Section \ref{analysis}. Finally, conclusions are drawn in Section \ref{conclusion}.

\section{Measurement Setup}\label{meas_setup}
The successful deployment of wireless communication systems requires a solid understanding and accurate modeling of wireless channel conditions (propagation characteristics) between the transmitter (Tx) and receiver (Rx). Channel sounding is a measurement technique used to gain that understanding.
\begin{figure*}[t]
\centering
\includegraphics[width=5.5in]{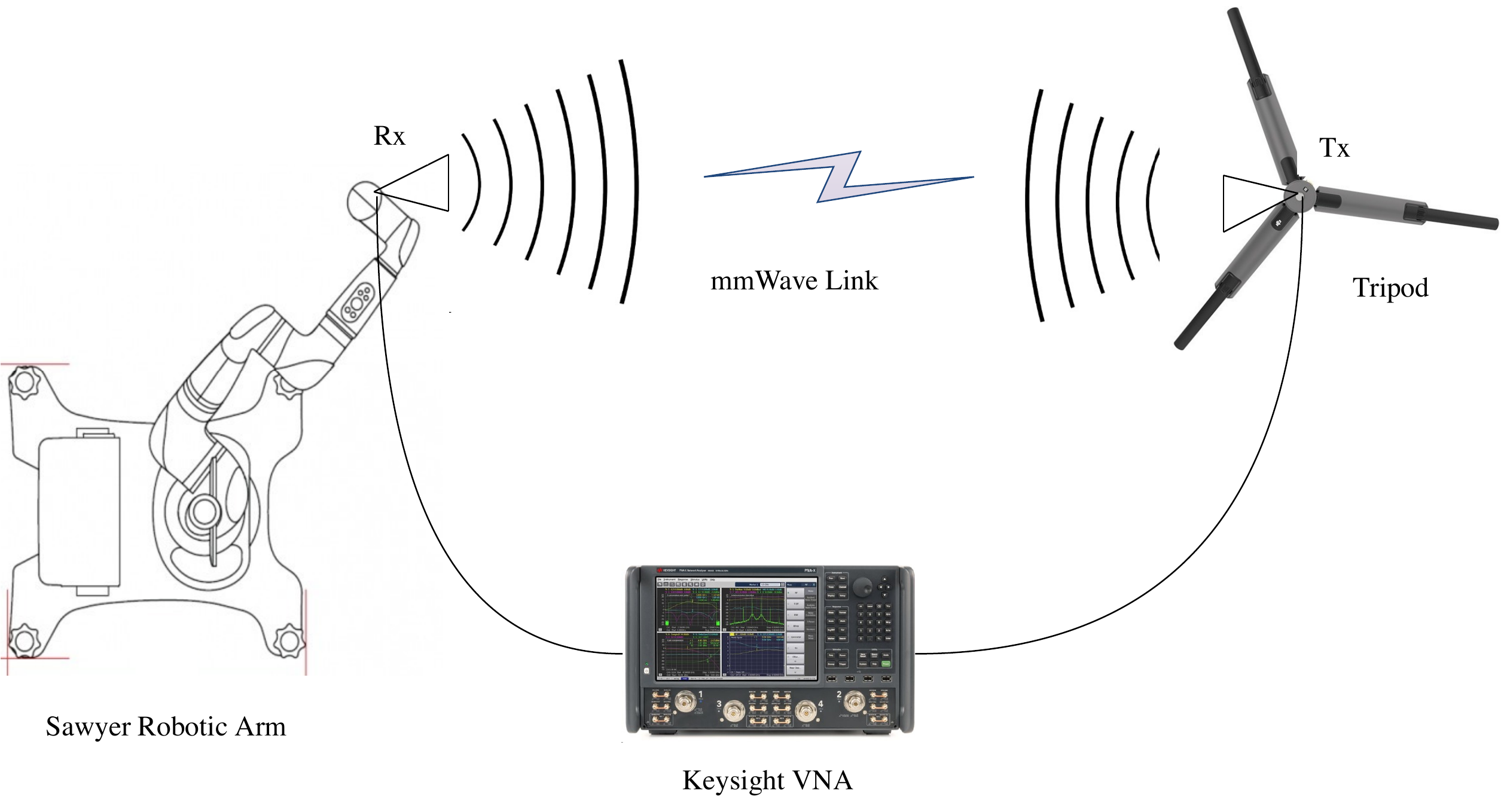}
 \caption{Top view of the measurement setup. The receiver is placed on the robotic arm and connected to VNA by a phase stable short cable, while  the transmitter is hooked on the tripod with a long cable. }
\label{measurement setup}
\end{figure*}
The channel sounding measurement method used in this study is based on continuous wave (CW) mode of VNA \cite{maccartney2017flexible,wu201528}. The Rx is hooked up on the robotic arm, while the  Tx is placed on a tripod. The Tx is positioned at different distances from the Rx  (3.5 ft to 23.5 ft) depending on measurements plan. The Tx is connected at port-1 of the VNA, while the Rx is connected at port-2. At each distance point (11 points in total, with 2 feet increments between the Tx and Rx), three samples of S21 parameters are recorded on the VNA. These multiple samples of S21 parameters are recorded to get a better redundant value in the Doppler spectrum calculations, and in channel path loss modeling. The measurement setup  is shown  in the Fig. \ref{measurement setup} with the details of all the measurement equipment  in Table \ref{measEquip}.

\setlength{\extrarowheight}{1pt}
\begin{table}[!ht]
\centering
\caption{The measurement equipment with the specifications.}
\label{measEquip}
\begin{tabular}{|c|c|}
\hline
\textbf{Equipment} & \textbf{Specifications} \\ \hline
\begin{tabular}[c]{@{}c@{}}Vector Network \\ Analyzer\end{tabular} & \begin{tabular}[c]{@{}c@{}}Keysight PNA-X (N5247A),\\ 10 MHz to 67 GHz\end{tabular} \\ \hline
Antenna (horn type) & \begin{tabular}[c]{@{}c@{}}Cernexwave (CRA28264015), \\ 26.5 GHz-40 GHz\\ Gain: 15 dBi, HPBW:18$^{\circ}$\end{tabular} \\ \hline
Waveguide transition & \begin{tabular}[c]{@{}c@{}}Cernexwave (CWK28264003F), WR-28, \\ Brass/Copper\end{tabular} \\ \hline
Cables & \begin{tabular}[c]{@{}c@{}}Fairview microwave (50 feet), \\ Mini-Circuits (5 feet)\end{tabular} \\ \hline
Robotic arm & \begin{tabular}[c]{@{}c@{}}Rethink Robotics (Sawyer), software: Intera,\\ 1 arm x 7 degree of freedom\end{tabular} \\ \hline
\end{tabular}
\end{table}

\begin{figure*}[t]
\begin{minipage}[t]{0.49\linewidth}
    \includegraphics[width=\linewidth,height=5cm]{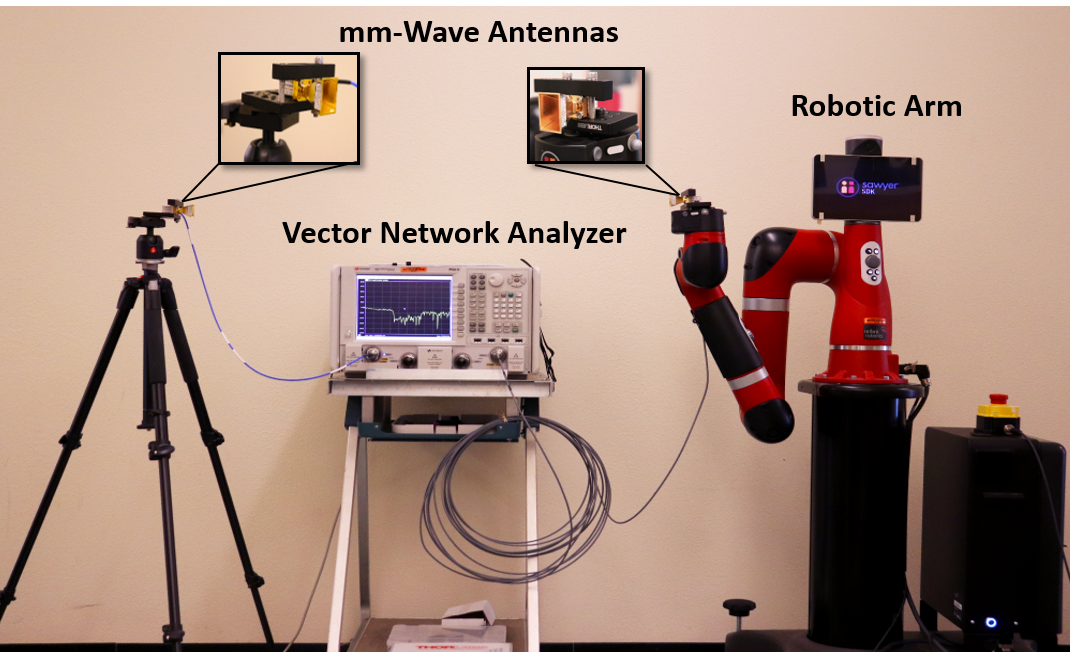}
    \subcaption{Equipment description.}
    \end{minipage}%
    \hfill%
\begin{minipage}[t]{0.49\linewidth}
    \includegraphics[width=\linewidth]{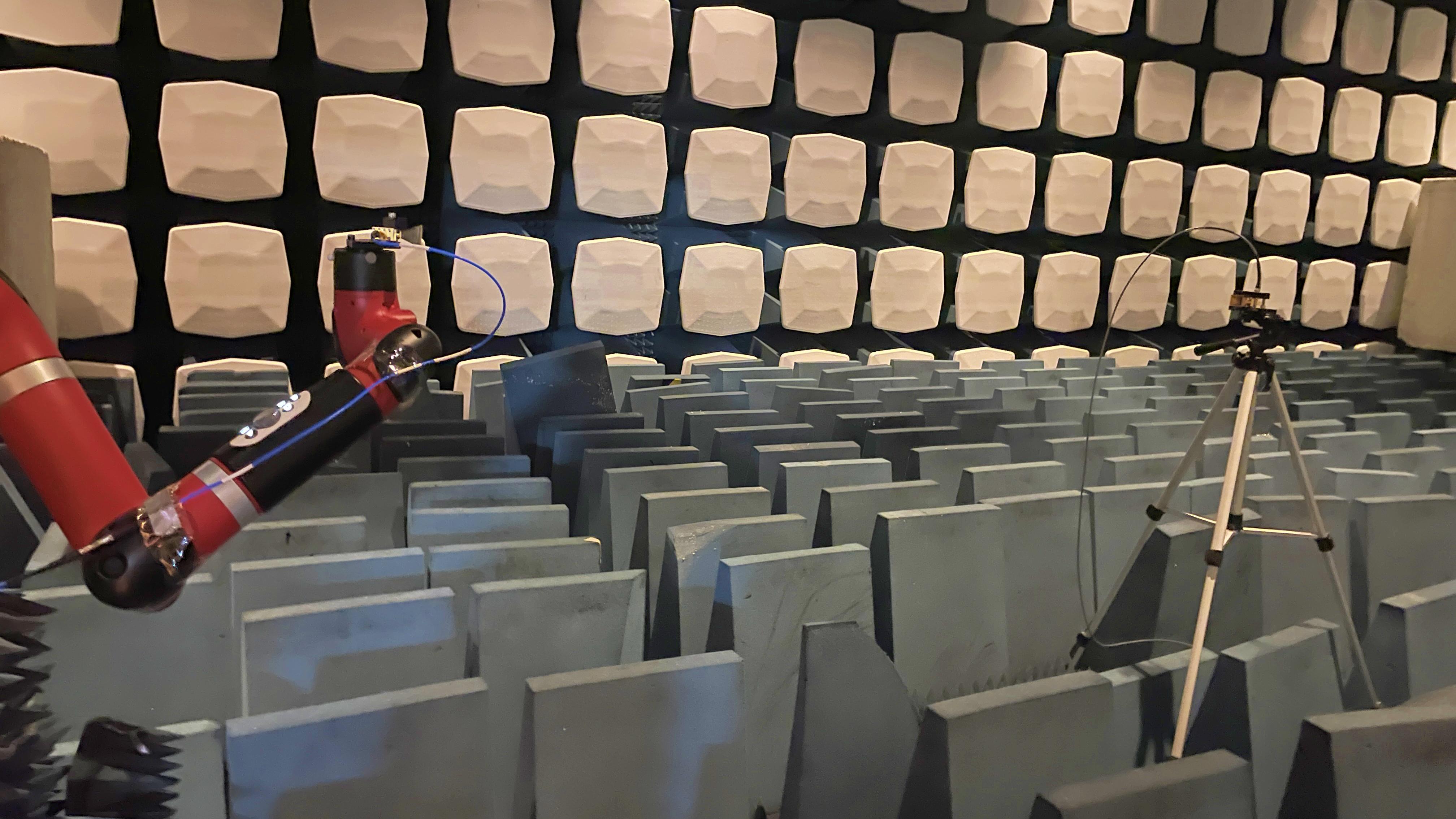}
    \subcaption{Actual setup in an anechoic chamber.}
  \end{minipage} 
\caption{Actual measurement setup with the transmitter on the tripod, and receiver hooked up at the robotic arm in an anechoic chamber.}
\label{real setup}
\end{figure*}

The VNA is operated at a frequency of 28 GHz with  the intermediate frequency (IF) bandwidth set at 300 Hz. This IF frequency is selected carefully in such a way that the expected  Doppler frequency can be covered within the selected IF  range. In addition, a total of 4096 points were selected in this CW (carrier wave) mode. It is worth to mention that increasing the IF bandwidth will definitely decrease the time required to capture the data at each distance for a given run, however, the scattering parameter data will suffer from higher noise floor as the noise floor level is linearly proportional to the IF bandwidth. Therefore, taking multiple readings with moderate IF bandwidth was a main factor in this measurement scenario.  With this selected 4096 sample points, the  sampling time is,  $Ts=4.4$ ms.  Thus, the Doppler range that could be captured with these settings will be in the range of $-fs/2=-114.49$ Hz to $+fs/2=+114.49$ Hz. 
Fig. \ref{real setup} shows the equipment, and the actual measurement  setup  in an anechoic chamber. The  Tx is at a height of 4 feet 4 inches on a tripod, while the  Rx is initially at the height of 4 feet 6 inches (at the beginning point)  on the robotic arm. The robotic arm is connected to a computer that executes a Python script to emulate the UAV motion with the wind turbulence model.  The corresponding scattering parameter data generated under such motion is captured on the VNA, and analyzed on a workstation later on. In the next section, we will dwell more into the details on the UAV motion emulation part with the robotic arm. 


\section{UAV Motion Emulation} \label{UAV_motion}

To emulate a real UAV motion under wind turbulence is not only a very challenging but also an interesting problem. This problem can be solved by using methods from robotics area, where the real UAV motion with the wind turbulence can be easily emulated by a robotic arm. The motion of this robotic arm is controlled by robotic operating system (ROS) from a computer. To create the  turbulence experienced by UAVs in atmosphere, a wind generation model is used.  

The  UAV motion is first simulated in MATLAB\textsuperscript \textregistered using a stochastic wind gust model (Dryden wind model) and a 6 degree-of-freedom (DOF) quadcopter dynamic model together with closed-loop controllers are used for hovering motion models. Then, the positions and altitude of the quadcopter generated from the simulations are emulated by the end-effector (end point) of the robotic arm. The simulation framework and the arm control is briefly discussed in the following subsections.

\subsection{Simulation of Quadcopter Motion with Wind}

We model the wind as a combination of a 3D mean wind $(\bar u,\bar v,\bar w)^\top$ and a 3D turbulence wind $(u,v,w)^\top$. The mean wind is specified in the north-east-down (NED) coordinates, while the turbulence wind is specified in another different frame (discussed later). In particular, the Dryden turbulence model~\cite{MIL-STD1990,MIL-HDBK2012} is used to generate turbulence. This  popular Dryden model has been extensively used for practical aviation applications (see e.g.,~\cite{beard_atmospheric_disturbances} for its application in small fixed-wing UAV). It is based on a stochastic formulation that incorporates knowledge of the energy spectrum of turbulence~\cite{von1948progress} and assumes a homogeneous frozen spatial turbulence. 
The Dryden turbulence model is implemented through  filtering operations on white noise signals.

\begin{figure}[!ht]
    \centering
    \includegraphics[width=1\linewidth]{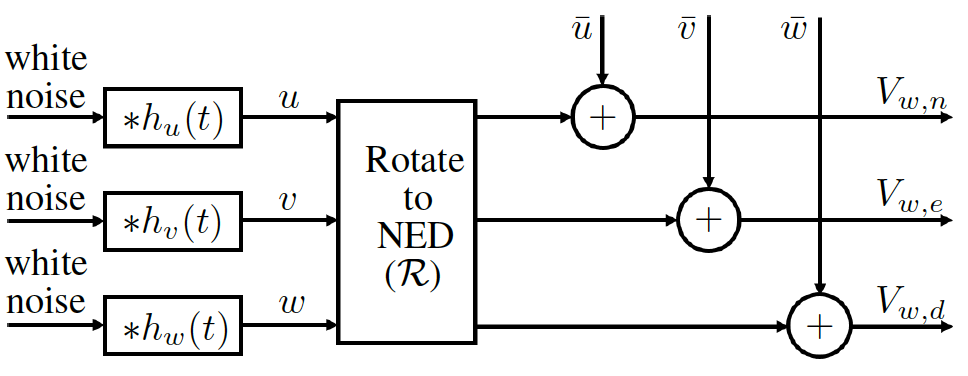}
    \caption{The Dryden wind is generated by combining the mean wind $(\bar u,\bar v,\bar w)^\top$ and the turbulence wind $(u,v,w)^\top$. $*h_\star(t)$ denotes the convolution with the impulse response of $H_\star(s)$ in~\eqref{eq:Hu}--\eqref{eq:Hw}, where $\star=u,v,w$.}
    \label{fig:dryden}
  \end{figure}

 Let the 3D turbulence velocity be $(u,v,w)^\top\in\mathbb{R}^3$. The $u$ component is aligned with the direction of the horizontal mean wind, i.e., $(\bar u,\bar v, 0)$, the $w$ component is aligned with the vertical (down) direction, and the $v$ component is aligned with the direction that completes a right-handed coordinate frame with the $v$ and $w$ directions. As shown in Fig.~\ref{fig:dryden}, the 3D velocities $u$, $v$, and $w$ in time domain are obtained by passing three independent white noise signals through three filters described by the following transfer functions~\cite[Section 4.4]{beard_atmospheric_disturbances} 
 \begin{equation}
     \begin{split}
    H_u(s)& =  \sigma_u\sqrt{\frac{2V_{a0}}{L_u}}\frac{1}{(s + \frac{V_{a0}}{L_u})},\label{eq:Hu}
     \end{split}
 \end{equation}
  \begin{equation}
     \begin{split}
H_{v}(s) &=  \sigma_{v}\sqrt{\frac{3V_{a0}}{L_{v}}}\frac{(s+\frac{V_{a0}}{\sqrt{3L_{v}}})}{(s + \frac{V_{a0}}{L_{v}})^2},\label{eq:Hv}
     \end{split}
 \end{equation}
\begin{equation}
     \begin{split}
H_{w}(s) &=  \sigma_{w}\sqrt{\frac{3V_{a0}}{L_{w}}}\frac{(s+\frac{V_{a0}}{\sqrt{3L_{w}}})}{(s + \frac{V_{a0}}{L_{w}})^2},\label{eq:Hw}
     \end{split}
 \end{equation}
respectively, where $ (\sigma^2_u, \sigma^2_v, \sigma^2_w)$ are the variances of the turbulence, $V_{a0}$ is an estimate of the quadrotor's airspeed, and $(L_u, L_v, L_w)$ are the turbulence length scales. In our experiments, we focus on horizontal, low turbulence effects and set $\sigma^2_u=0.53$, $\sigma^2_v=0.53$, and $\sigma^2_w=0$. For low altitude, $L_w$ is set to the altitude in feet.

In our experiments, we set $(L_u, L_v, L_w)$ to $(200,200,50)$ ft. Since the quadcopter is controlled in the hover mode, we set $V_{a0}$ to be the mean wind speed, i.e., $\sqrt{\bar u^2+\bar v^2+\bar w^2}$, where $\bar u = 2$ m/s, $\bar v = -1$ m/s, and $\bar w = 0$ m/s.  The output of the three filters is a time-series representation of the turbulence, which is further combined with the non-zero mean wind $(\bar u,\bar v,\bar w)^\top\in\mathbb{R}^3$. The resulting 3D wind in the NED frame is given by

\begin{equation}
    V_w(t) = (\bar u,\bar v,\bar w)^\top + \mathcal{R}(u(t),v(t),w(t))^\top,
\end{equation}
where $\mathcal{R}\in\mathbb{R}^{3\times 3}$ is a 3D rotational matrix that converts $(u(t),v(t),w(t))^\top$ to the NED frame. The process of generating the Dryden wind is shown in Fig.~\ref{fig:dryden} below.  

\begin{figure}[t]
\centering
\includegraphics[width=1 \linewidth]{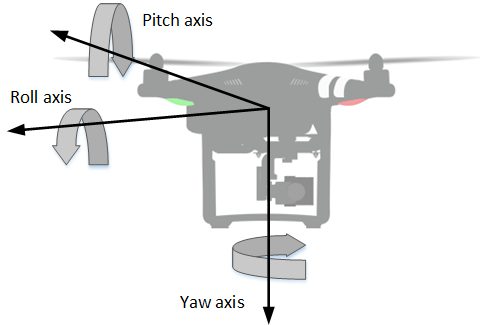}
 \caption{Pitch, roll and yaw axis of a UAV. }
\label{yaw}
\end{figure}

We use a standard formulation of the quadcopter dynamics~\cite{Beard1}, with the addition of a nonlinear drag term $f_d\in\mathbb{R}^3$ to model the wind effect. The dynamics of the quadcopter is given by,

\begin{align}
\begin{bmatrix}
\ddot p_n \\ \smallskip
\ddot p_e \\ \smallskip
\ddot p_d \\ \smallskip
\dot \phi \\ \smallskip
\dot \theta \\ \smallskip
\dot \psi \\\smallskip
\dot \omega_p \\ \smallskip
\dot \omega_q \\ \smallskip
\dot \omega_r \\
\end{bmatrix}
=
\begin{bmatrix}
(-\cos\phi \sin\theta \cos\psi - \sin\phi \sin\psi) \frac {F} {m} + \frac{f_{d,n}}{m} \\
\smallskip
(-\cos\phi \sin\theta \sin\psi + \sin\phi \cos\psi) \frac {F} {m} + \frac{f_{d,e}}{m} \\
\smallskip
g - (\cos\phi \cos\theta) \frac {F} {m} + \frac{f_{d,d}}{m} \\
\smallskip
\omega_p + \sin\phi\tan\theta\,\omega_q + \cos\phi\tan\theta\,\omega_r\\
\smallskip
\cos\phi\,\omega_q-\sin\phi\,\omega_r\\
\smallskip
\frac{\sin\phi}{\cos\theta}~\omega_q+\frac{\cos\phi}{\cos\theta}~\omega_r\\
\smallskip
\frac{J_y - J_z}{J_x} \omega_q\omega_r+  \frac 1 {J_x} \tau_\phi \\ \smallskip
\frac{J_z - J_x}{J_y} \omega_p\omega_r \dot \psi + \frac 1 {J_y} \tau_\theta \\
\smallskip
\frac{J_x - J_y}{J_z} \omega_p\omega_q + \frac 1 {J_z} \tau_\psi \\
\end{bmatrix},\label{eq:dyn}
\end{align}
where $m$ and $(J_x,J_y,J_z)$ are the mass and the moment of inertia of the quadcopter, respectively, $(p_n, p_e, p_d)$ are the north, east, and down positions in the inertial frame, $(\phi, \theta, \psi)$ are the roll, pitch, and yaw angles, respectively, $(\omega_p,\omega_q,\omega_r)$ is the angular velocity in the body frame, $(F, \tau_\phi, \tau_\theta, \tau_\psi)$ are the force and the moments in the designated directions, and $f_d = (f_{d,n}, f_{d,e}, f_{d,d})^T$ is the drag force in the north, east and down directions. The pitch, roll, and yaw axis are also shown in Fig. \ref{yaw}. The drag force $f_d$ is modeled in a quadratic form as $f_d =  C_d(V_w - \dot{p})|V_w - \dot{p}|$, where $\dot p$ is the ground velocity of the quadcopter equal to $(\dot p_n, \dot p_e, \dot p_d)$ and $C_d$ is a drag coefficient matrix. 

In the quadcopter dynamics, we have also included motor dynamics and several secondary aerodynamic effects, such as blade flapping and air-relative velocity effects. Closed-loop PID (Proportional-Integral-Derivative) Controllers are designed to achieve tracking of a given waypoint. We refer interested readers to \cite{allison2019estimating,allison2019wind} for details on the quadcopter dynamics and the controller design.

\subsection{Emulation of Quadcopter Motion using the Sawyer Robotic Arm} 

The MATLAB simulator generates NED positions, and the yaw, pitch, roll (YPR) orientations, as well as their time derivatives, at a rate of 10 Hz. These state variables are formatted and dumped to a raw text file. A Python program reads these NED and YPR coordinates and uses them to generate a series of time-dependent waypoints for the Sawyer robotic arm. The Intera software development kit (SDK) is used as an interface between the desired waypoints and the ROS messages needed to command the Sawyer robotic arm. For this application, the Intera Motion Controller Interface is implemented in a joint control mode. This mode generates trajectories based on a series of joint commands. The native inverse kinematics solver is used to convert waypoints from the end-effector frame to a given set of joint positions.

\subsection{Emulation of Linear Motion using the Sawyer Robotic Arm}
The MATLAB software is  used to generate the waypoints from the Dryden motion generation profile.  In order to precisely control the velocity, the Joint Velocity Control mode was implemented in the Intera SDK. This contol mode accepts velocity commands in the joint frames, not the end-effector frame, and as such the Jacobian matrix \textbf{$J(\theta_1, \theta_2, ...,  \theta_7)$} is needed to translate commanded velocities between frames. Specifically, the psuedo-inverse Jacobian matrix is needed to convert velocities from the end-effector frame to the joint frames as follows,
\begin{align}
J(\theta_1, \theta_2, ..., \theta_7) = 
\begin{bmatrix}
\frac{\partial x}{\partial \theta_1} & \frac{\partial x}{\partial \theta_2} & ... & \frac{\partial x}{\partial \theta_7}  \\ 
\smallskip
\frac{\partial y}{\partial \theta_1} & \frac{\partial y}{\partial \theta_2} & ... & \frac{\partial y}{\partial \theta_7} \\
\smallskip
\frac{\partial z}{\partial \theta_1} & \frac{\partial z}{\partial \theta_2} & ... & \frac{\partial z}{\partial \theta_7} \\
\smallskip
\frac{\partial \phi}{\partial \theta_1} & \frac{\partial \phi}{\partial \theta_2} & ... & \frac{\partial \phi}{\partial \theta_7} \\
\smallskip
\frac{\partial \theta}{\partial \theta_1} & \frac{\partial \theta}{\partial \theta_2} & ... & \frac{\partial \theta}{\partial \theta_7} \\
\smallskip
\frac{\partial \psi}{\partial \theta_1} & \frac{\partial \psi}{\partial \theta_2} & ... & \frac{\partial \psi}{\partial \theta_7} 
\end{bmatrix},\label{eq:jac}
\end{align}

\begin{align}
Q = J^{-1} V,\label{eq:jac_inv}
\end{align}
where $J^{-1}$ is a $7\times 6$ matrix representing the psuedo-inverse of the Jacobian, $V$ is a $6\times 1$ matrix representing the linear and angular velocities of the end-effector, and $Q$ is a $7\times 1$ matrix representing the joint velocities.

\begin{figure}[t]
\centering
\includegraphics[width=1 \linewidth]{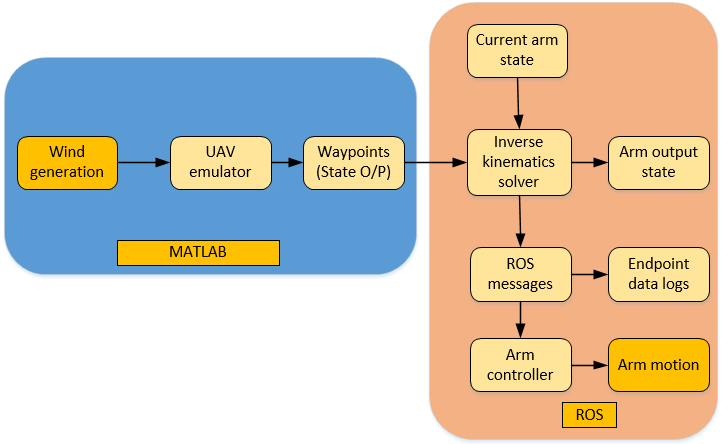}
 \caption{Arm motion dynamics. }
\label{armMotion}
\end{figure}

This method provides precise control of the arm's velocity at speeds under 0.5 m/s. However, drift was introduced in the x-axis as the arm reached higher speeds and the end-effector reached its outer boundaries. As such, a P (proportional) controller was used to minimize the drift along the x-axis.  This whole UAV motion emulation dynamics\footnote{Note that  this motion doesn't include the propeller motion which causes amplitude modulation in the received signal.} is summarized  in Fig. \ref{armMotion}. An actual snapshot of the UAV motion can also be visualized in this GitHub repository\footnote{\url{https://github.com/amitkac/uavSim}}. Once the robotic arm  is emulating the real UAV motion under turbulence, the next step is to measure scattering parameters from the VNA. 

\subsection{De-embedding the Phase Changes Caused by Cable Movement}
One important point to consider is that because of this robotic arm motion, the cable connected to VNA and antenna on the arm will also move, and this will result in phase changes at the received signal. To take care of such scenario, it is therefore very critical to utilize a phase stable cable, and also the cable effects need to be taken out from the final measured data. The de-embedding \cite{bauer1974embedding,keysight} of phase can be performed by first determining two-port S-parameters of the cable  from the measurements with short, open, and load standards while doing the same motion. The process is similar to ``Adapter Characterization" function  within Keysight VNAs \cite{keysight}, which involves Short-Open-Load calibrations at two different reference planes. The only difference is that this process is performed over time-swept data rather than frequency-swept data. 
\begin{figure}[t]
    \centering
    \includegraphics[width=1\linewidth]{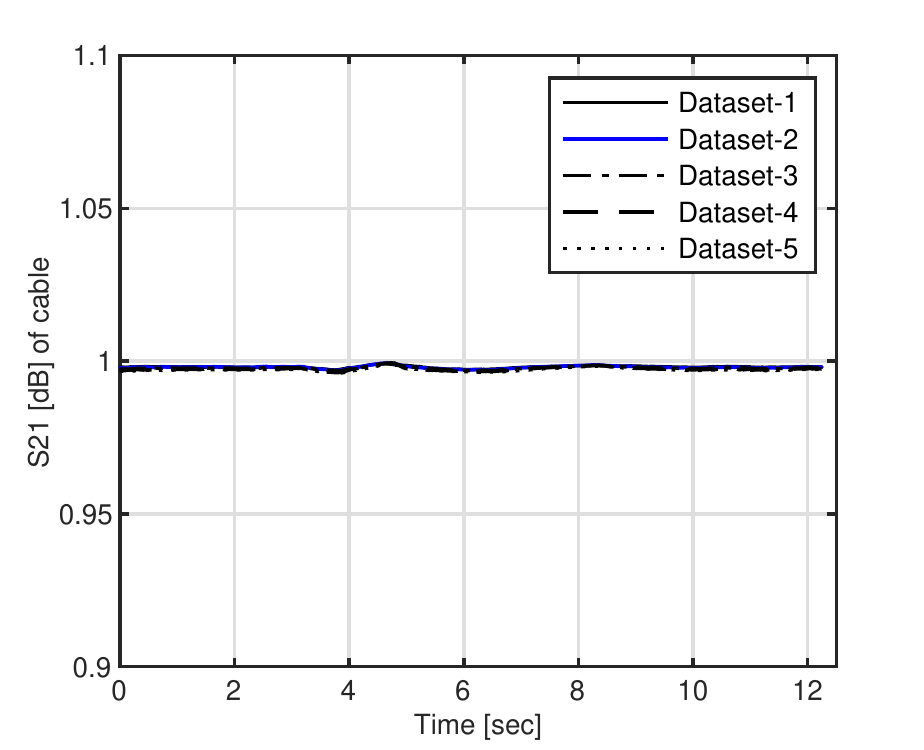}
    \caption{Magnitude variations of S21 data collected in the repeatablity test. }
    \label{mag5}
\end{figure}

\begin{figure}[t]
    \centering
    \includegraphics[width=1\linewidth]{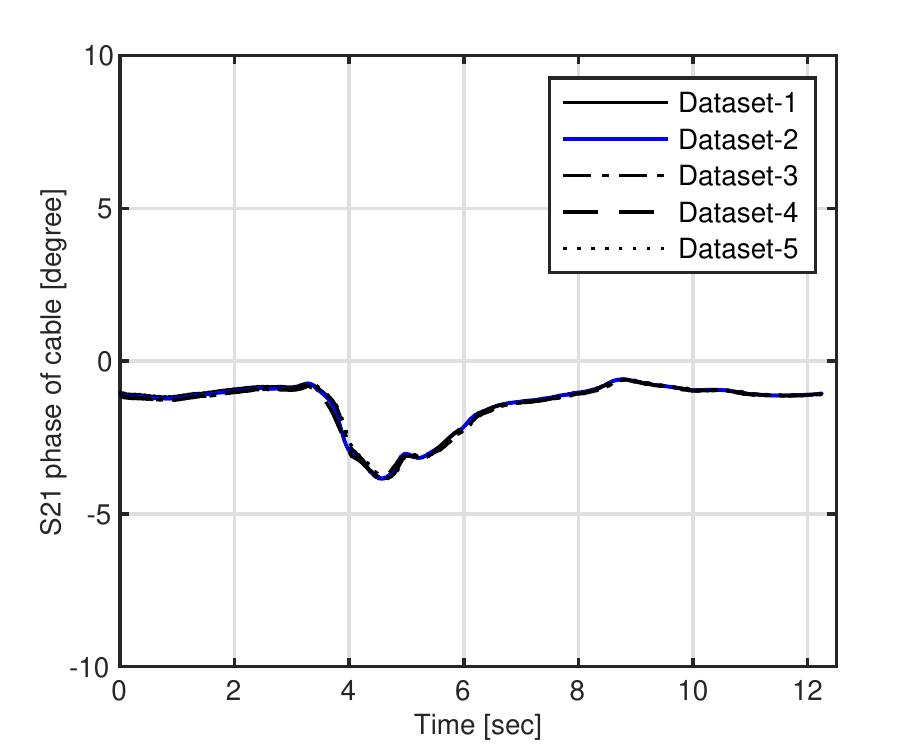}
    \caption{Phase variations of the S21 data collected in the repeatablity test.}
    \label{phase5}
\end{figure}

A repeatability test (5 times with same motion) was also performed to check the variance in the S-parameters. In our case,  it was found that the phase variation (min-max) after de-embedding  ranged from -4 degree to +0.6 degree, while the magnitude variation (min-max) ranged from   -0.033 dB to -0.007 dB for the phase stable cable in an  anechoic chamber. Fig. \ref{mag5} and Fig. \ref{phase5} show these variations in our scenario. Later on, offset calibration was also done by removing S21 of the cable from the measured data.

In the next section, the signal processing of this processed data, and  the corresponding analysis to determine the Doppler spread and path loss modeling will be presented in detail.

\section{Analysis of Doppler Spread and Path Loss Modeling} \label{analysis}

In this section, first the data processing of the pre-processed (de-embedded and offset-calibration) VNA data (S-parameters), and its correlation with the motion profile extracted from the ROS logs will be analyzed, and then the analysis of Doppler spread and path loss will be discussed in detail. Let the transmitted signal, $ x(t)=\Re\{ b(t) e^{j2\pi f_c t}\}$, where $b(t)=x_I(t) + jx_Q(t)$ is a complex baseband signal with in-phase and quadrature components as $x_I(t)$ and $x_Q(t)$, and $f_c$ is the carrier frequency. The signal $b(t)$ is also known as the complex signal envelope or an equivalent low pass signal of $x(t)$.  Ignoring the noise in the system, the corresponding received signal with the line of sight (LOS), and all resolvable multipath is given as \cite{goldsmith2005wireless},
\begin{equation}
    \begin{split}
        y(t)&=  \Re \bigg\{ \mathlarger{\sum}_{n=0}^{N(t)} \alpha_n(t) b(t-\tau_n(t))e^{j2\pi[f_c(t-\tau_{n}(t)) + \phi_{d_n} +\phi_0] } \bigg\},
                    \end{split}
\end{equation}

\begin{equation}
    \begin{split}
        y(t) &=  \Re \bigg\{ \bigg[ \mathlarger{\sum}_{n=0}^{N(t)} \alpha_n(t) b(t-\tau_n(t))e^{-j(2\pi f_c \tau_{n}(t) +  \phi_{d_n} +\phi_0)} \bigg]\\
        &~~~~~~~~~~~~~~~~~~~~\times e^{j2\pi f_c(t)} \bigg\}.\\
            \end{split}
\end{equation}
which can be further written as,
\begin{equation}
    \begin{split}
y(t)&=\Re \bigg\{ \bigg[ \int_{-\infty}^{+\infty} h(t,\tau) b(t-\tau) d\tau \bigg] e^{j2\pi f_c(t)} \bigg\},
\end{split}
\end{equation}
where $h(t,\tau)$ is the channel impulse response, 
\begin{equation}
    \begin{split}
        h(t,\tau)&=\mathlarger{\sum}_{n=0}^{N(t)} \alpha_n(t) e^{-j(2\pi f_c \tau_{n}(t) +  \phi_{d_n} +\phi_0)} \delta(\tau -\tau_n),
    \end{split}
\end{equation}
where $\alpha_n(t)$ is a function of path loss and shadowing, $N(t)$ is the $N$-th resolvable multipath, $\tau_n$ is the $N$-th path delay, $\phi_0$ is the phase offset, and $\phi_{d_n}$ is the  Doppler phase shift of this $N$-th path. Now, when the Tx or Rx is moving, the change in the distance  over a short time interval $\Delta t$ will cause the  phase to change as $ \phi_{d_n} \approx 2 \pi (v/\lambda) \Delta t \cos \theta$, where $\theta$ is the arrival angle of the received signal relative to the direction of motion, and $v$ is the velocity of receiver towards transmitter in the direction of motion. Therefore, the corresponding Doppler frequency will be then  given as,
\begin{equation}
\begin{split}
      f_D&=\frac{1}{2\pi}\frac{\phi_{d_n}}{\Delta t}=\frac{v}{\lambda} \cos{\theta}, 
\end{split}
\end{equation}
where $\lambda$ is the wavelength of the signal.  The data collected at VNA during the measurement campaign are the S-parameters, also known as scattering parameters, over time (CW mode).  


Since the Tx is connected at port-1 and Rx at port-2, therefore,  the S21 parameter will represent the channel transfer function (channel power). These S-parameters  depend on time, distance and frequency.  Thus,  S21 parameter as a channel transfer function of time, distance, and frequency can be written as\cite{kachroo2019unmanned,pozar2011microwave},
\begin{equation} \label{s21fxn}
    \begin{split}
        \text{S21}_{dB}(t,f,d)&=20\log_{10}(|H(t,f,d)|)
    \end{split}
\end{equation}

As the VNA is set in CW time mode with frequency kept fixed at 28 GHz, the only variable in \eqref{s21fxn} would be the discrete time  points at a given distance. Therefore, at a given distance $d$, \eqref{s21fxn} can be rewritten as,
\begin{equation}
    \begin{split}
        \text{S21}_{dB}(t,f,d)&=\text{S21}_{dB}(t)\\
    \end{split}
\end{equation}


\begin{figure}[t]
\includegraphics[width=1\linewidth]{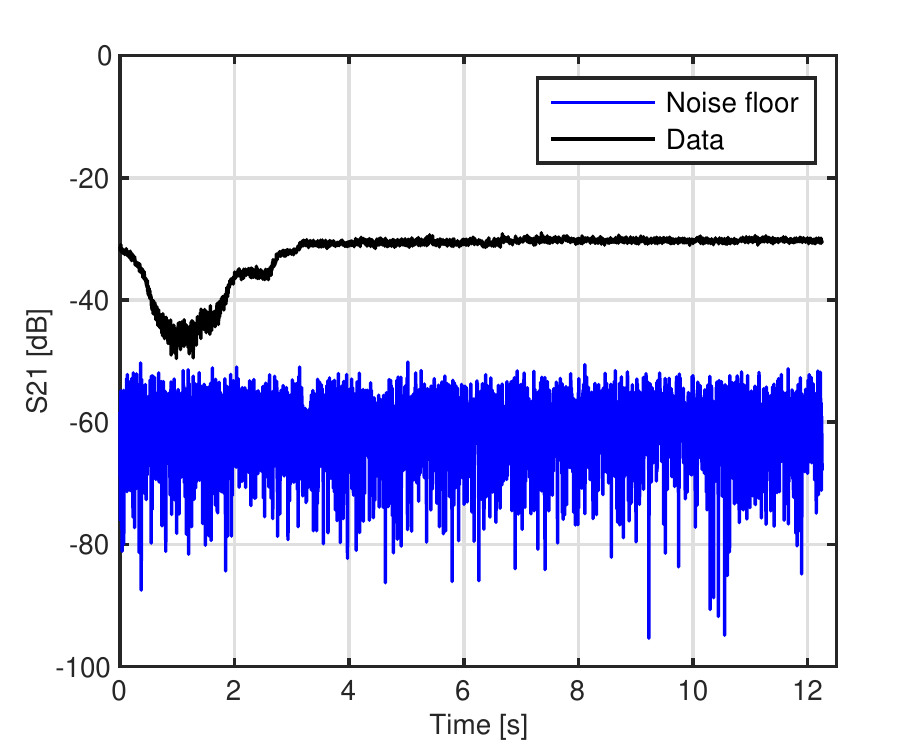}
\caption{Measurement data with  noise floor at 3.5 feet over time.}
\label{mag}
\end{figure}

\begin{figure}[t]
\includegraphics[width=1\linewidth]{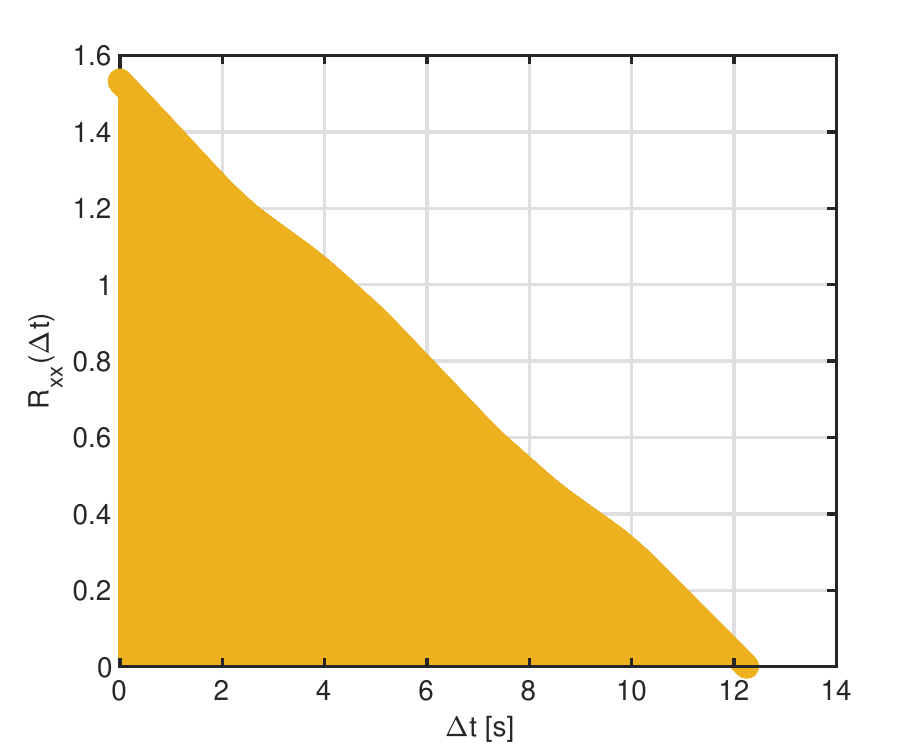}
\caption{One sided autocorrelation of S21 data at 3.5 feet over delay.}
\label{phase}
\end{figure}
These discrete points of S21 parameter can be written in a complex vector form as, $x[n]=\text{S21}_{dB}(t)\delta(t-t_i)$, where $t_i$ is the discrete time point, $\delta(\cdot)$ is the delta function, and the index  $i$ goes from 0 to 4095. Because, the S21 parameters are complex numbers, therefore, the frequency response after taking the fast Fourier transform (FFT) would  also be   complex in nature. Therefore, 
\begin{equation}
    \begin{split}\label{fft}
       X[k]= \frac{1}{N} \mathlarger{\sum}_{n=0}^{N-1}x[n]e^{-j 2\pi k n/N },
    \end{split}
\end{equation}
where $k$ is the point in the frequency domain, $n$ is the point in time domain, and $x[n]$ is the  discrete time domain input (vector of all 4096 points). Fig . \ref{mag} shows a sample S21 data with noise floor collected at 3.5 feet.

Moreover, the UAV motion in general, and the current emulated motion of the  UAV by the robotic arm  is  very dynamic in nature. The motion not only considers  the modelled wind gust (Dryden wind model) in the atmosphere but also considers the UAV structure. For a process to be assumed wide sense stationary with uncorrelated scattering (WSSUS)\cite{khuwaja2018survey,bello1973aeronautical,matolak2012air,walter2011doppler}, the mean should be constant over time, and the autocorrelation function should be independent of the time. On closer inspection of the measured data (Fig . \ref{mag} and Fig. \ref{phase}), it was observed that this condition can be assumed true in anechoic chamber settings\footnote{In real scenario of outdoor or indoor environment, the WSSUS conditions may not be true.}.


\subsection{Doppler Spread}

\begin{figure}[t]
\includegraphics[width=1\linewidth]{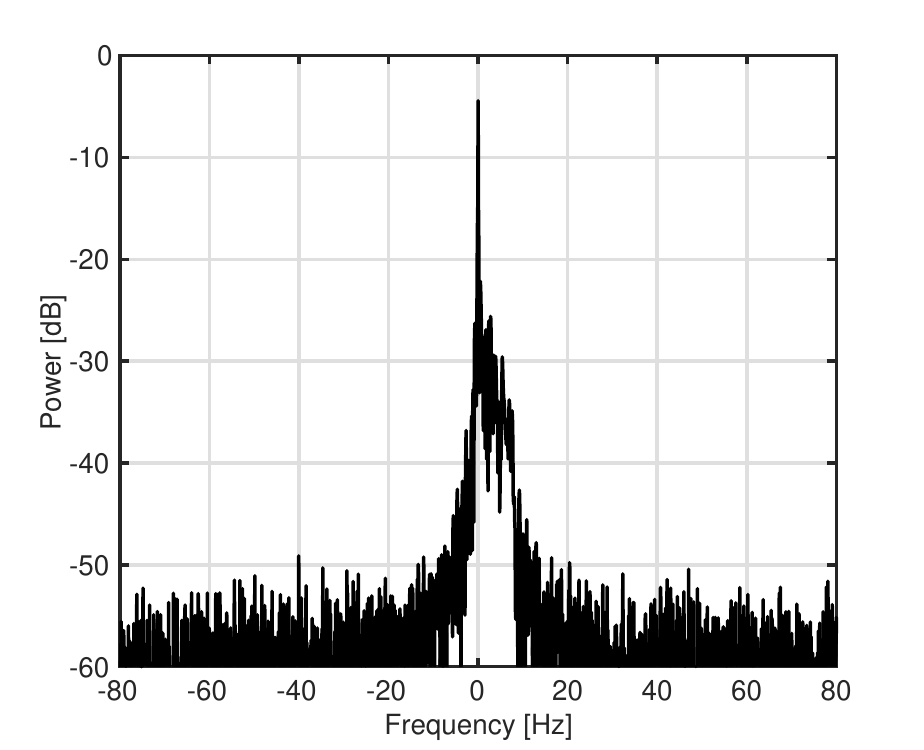}
\caption{Doppler spectrum at 5.5 feet.}
\label{doppler101}
\end{figure}

As mentioned, in our set up, 4096 points were set in CW time mode, this number was chosen so that the Discrete Fourier transform (DFT) can be efficiently calculated by using  the FFT algorithm as the data points will be in the form of  $2^n$. As $N$=4096,  \eqref{fft} can be rewritten as follows,
\begin{equation}
    \begin{split}
       X[k]= \frac{1}{4096} \mathlarger{\sum}_{n=0}^{4095}x[n]e^{-j 2\pi k n/4096 },
    \end{split}
\end{equation}
where $k=0,1,\dots 4095$. Doppler spectrum after taking the FFT of S21 parameters at a distance of  5.5 feet is shown  in Fig. \ref{doppler101}. 
\begin{figure}[t]
    \includegraphics[width=1\linewidth]{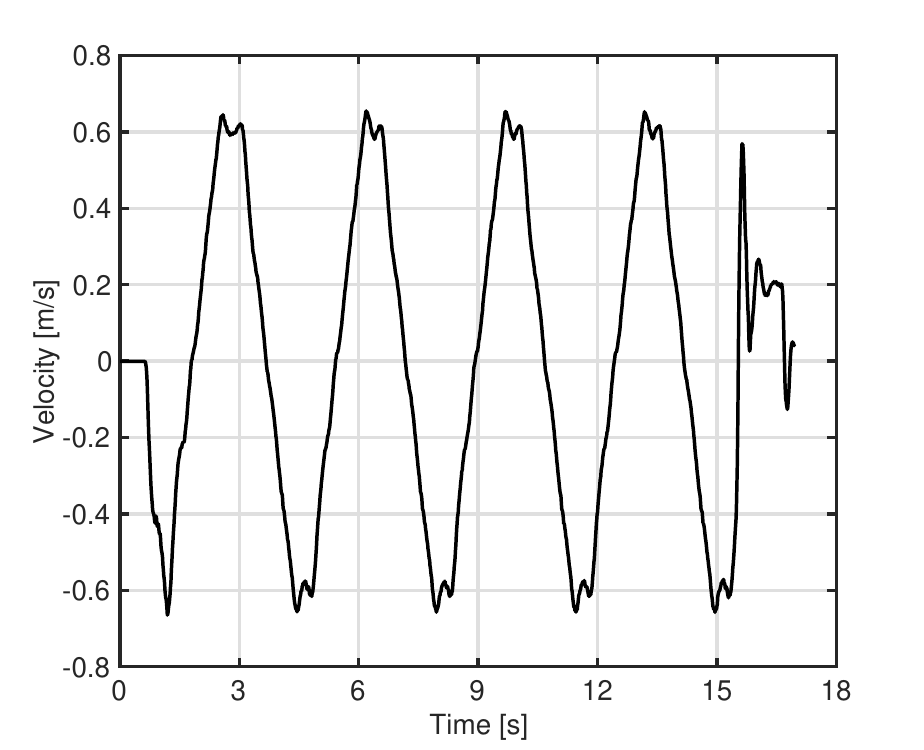}
    \caption{Controlled linear change of the arm speed in an anechoic chamber.}
    \label{velocity}
\end{figure}
To demonstrate the accuracy of the collected data for the Doppler analysis, we reprogrammed the robotic arm to do a controlled motion (to and from) with a pre-defined linear speed in an anechoic chamber. The robotic operating system, also known as ROS, captures various fields of the end point (velocity and displacements) of the arm in the system logs  in  \textit{.bag} format. These files are later  converted to \textit{.csv}  format to process with the MATLAB software.
\begin{figure}[t]
    \includegraphics[width=1\linewidth]{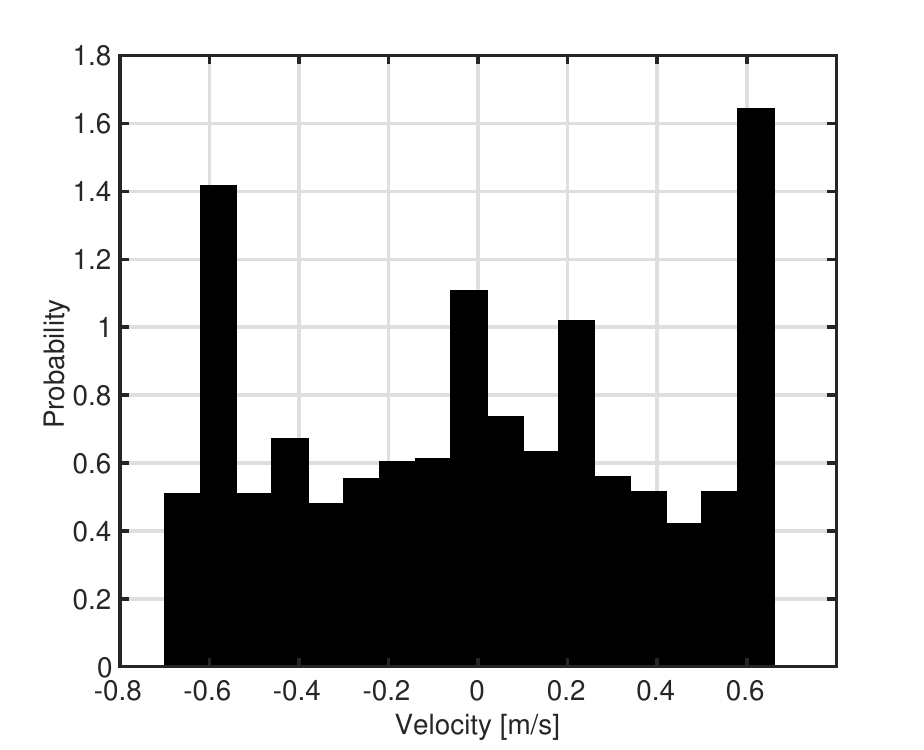}
    \caption{Velocity distribution over time.}
    \label{velocityD}
  \end{figure}

  \begin{figure}[t]
    \includegraphics[width=1\linewidth]{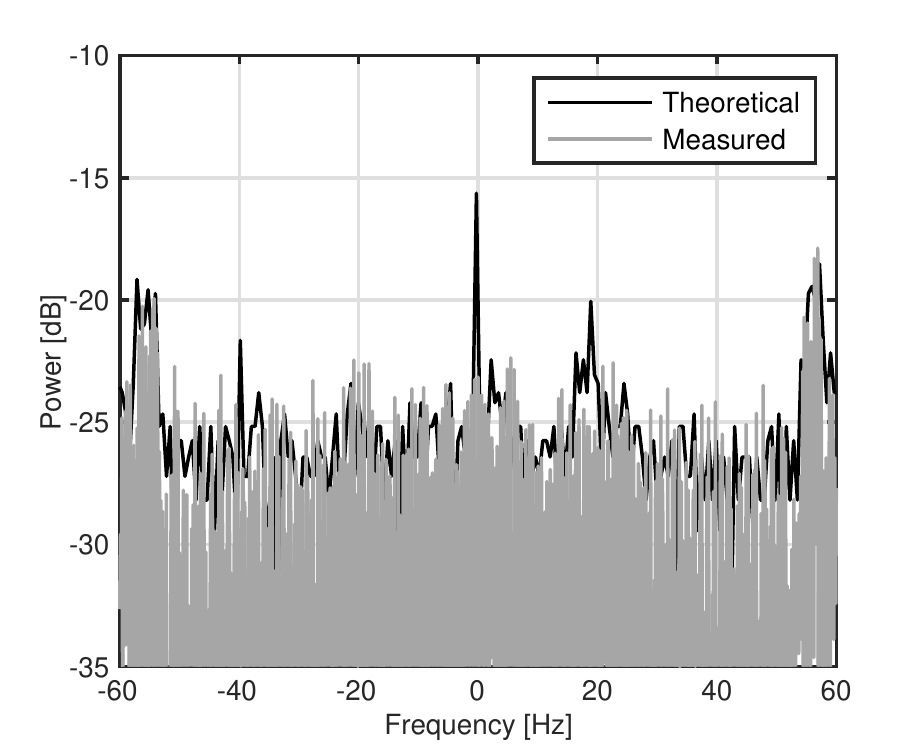}
    \caption{Theoretical Doppler with Measured Doppler at 5.5 feet.}
    \label{onTop}
\end{figure}

On close analysis, it can be  observed that the velocity of this control system will go from 0 m/s to 0.58-0.63 m/s and will stay at this peak there for some time in one direction and the same while going back (0 m/s to -0.58 to -0.61 m/s).  The corresponding probability density function (pdf) of  velocity change over time (Fig. \ref{velocity}) is shown in Fig. \ref{velocityD}. Intuitively, since velocity is proportional to the ideal Doppler observed, the  frequency corresponding to the peaks in the velocity pdf will also be seen in the Doppler spectrum.  For example, speed at the peak of 0.59 m/s to 0.64 m/s will correspond to a theoretical Doppler of 55.1 Hz to 59.7 Hz, while the speed of  -0.58 m/s to -0.64 m/s will correspond to Doppler of  -54.17 Hz  and -59.7 Hz, which can be seen in the Doppler spectrum\footnote{The theoretical Doppler power is offseted by 4 dB to fit the measured Doppler spectrum plot.} as shown in Fig. \ref{onTop}.   

\begin{figure}[t]
\includegraphics[width=1\columnwidth]{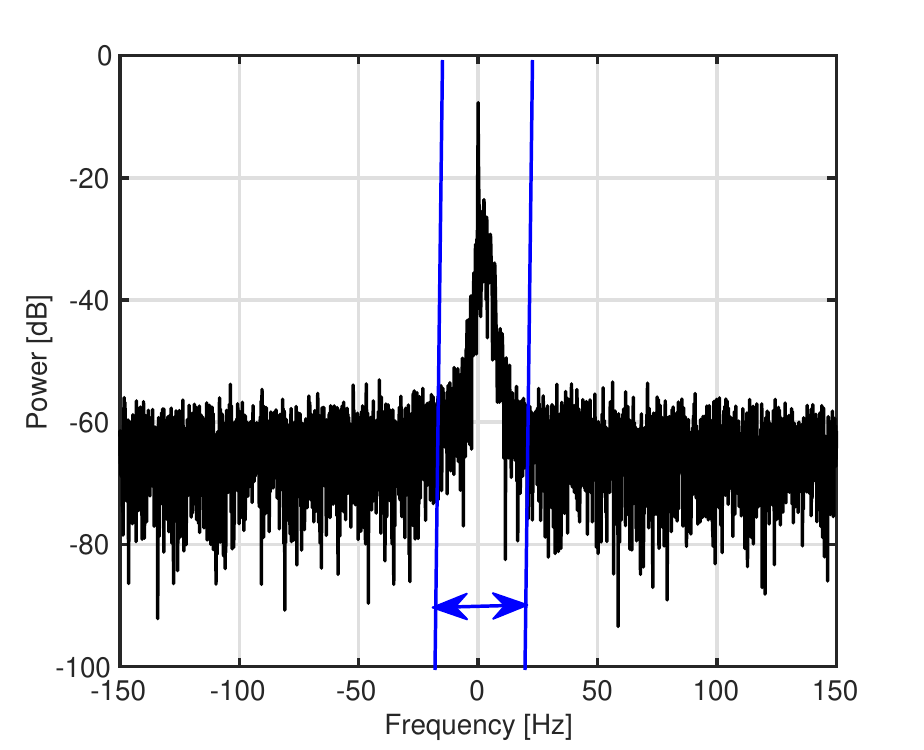}
\caption{Doppler spread at 11.5 feet in the Anechoic chamber environment.}
\label{anechoicCh}
\end{figure}

\setlength{\extrarowheight}{1pt}
\begin{table}[t]
\centering 
\caption{Maximum Doppler spread with distance in Anechoic chamber.}
\label{doppler22}
\begin{tabular}{|c|c|c|c|}
\hline
\multicolumn{1}{|l|}{Noise floor} & Distance [feet] & -ve freq. {[}Hz{]} & +ve freq.{[}Hz{]} \\ \hline
\multirow{11}{*}{-60 dB} & 3.5 & -25.3844 & 24.9763 \\ \cline{2-4} 
 & 5.5 & -20.5687 & 20.3328 \\ \cline{2-4} 
 & 7.5 & -18.5282 & 21.14 \\ \cline{2-4} 
 & 9.5 & -26.0374 & 19.9157 \\ \cline{2-4} 
 & 11.5 & -16.4876 & 22.446 \\ \cline{2-4}
 & 13.5 & -20.3238 & 20.6503 \\ \cline{2-4}
 & 15.5 & -20.079 & 20.5687 \\ \cline{2-4}
 & 17.7 & -18.5282 & 19.5076 \\ \cline{2-4}
 & 19.5 & -16.0795 & 17.5487 \\ \cline{2-4}
 & 21.5 & -17.4671 & 20.6503 \\ \cline{2-4}
 & 23.5 & -20.6503 & 20.4871 \\
  \hline
\multicolumn{2}{|l|}{Average Doppler [Hz]} & -19.4750 &	20.3204\\ \hline
\end{tabular}
\end{table}

The resulting maximum  positive Doppler spread (motion towards the receiver) and negative Doppler spread (motion away from the receiver) at each distance in an Anechoic chamber is shown in Table \ref{doppler22} with respect to the noise floor of -60 dB. The average Doppler spread in the Anechoic chamber was found to be around  +20 Hz to -20 Hz. Fig. \ref{anechoicCh} shows an example of the Doppler spectrum at 11.5 feet.  These Doppler spread parameters are paramount in designing the symbol duration in a wireless communication system. Specifically, the coherence time depends on this Doppler spread, and if not properly designed for it, will result in frequency selective fading.


\subsection{Path Loss Modelling}

\begin{figure}[t]
\centering
\includegraphics[width=1 \linewidth]{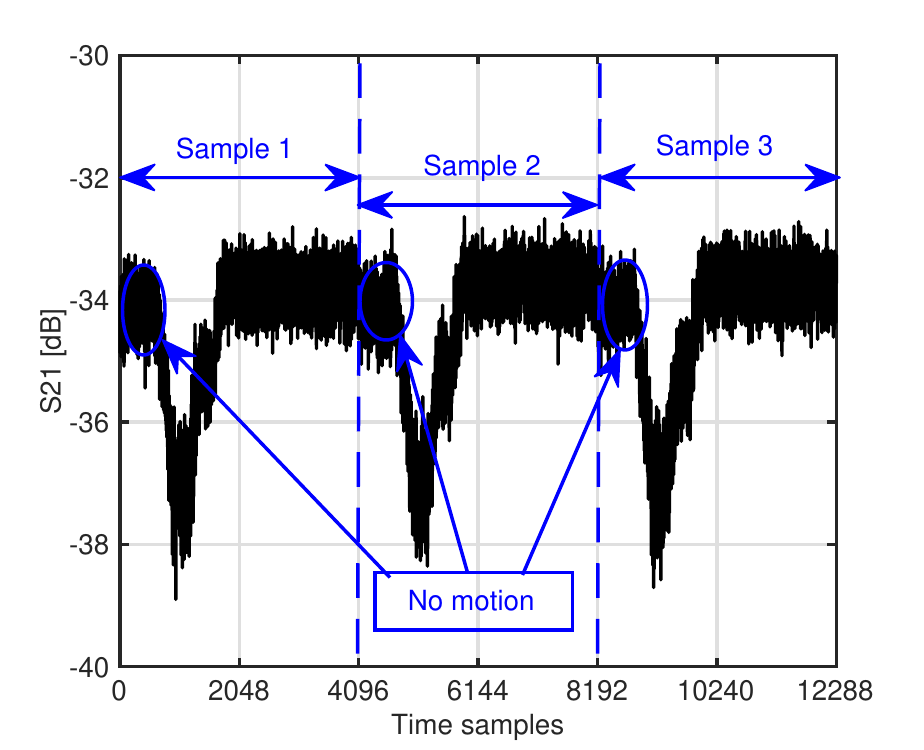}
 \caption{No motion portion in the appended data at 3.5 feet. }
\label{AmpStat}
\end{figure}

The next part is to statistically model the path loss characteristics of the wireless channel\footnote{Note that in this work, the effect of variation of elevation, and elevation angle on the Doppler is not considered.}. As mentioned in earlier section, at each distance point, 3 samples of scattering data were recorded. An example of such 3 samples of 4096 data points appended with each other is shown in Fig. \ref{AmpStat}. The first initial points (highlighted in Fig. \ref{AmpStat}) in these S21 power samples represent no-motion or idle motion of UAV. 
These points can be filtered from the appended samples  for each distance, and therefore, can be utilized to determine the distance dependent path loss exponent value in an anechoic environment.

The path gain as a function of distance can be modelled as\cite{goldsmith2005wireless,kachroo2019unmanned},
\begin{equation}
    \begin{split}
        \text{PG}_{dB}(d) &=  \text{PG}_{dB}(d_0)-10n \log_{10} \bigg( \frac{d}{d_0}\bigg) +X_\sigma (d),
    \end{split}
\end{equation}
where $\text{PG}_{dB}(d_0)$ is the path gain at a reference distance $d_0$, which is 1 meter in our scenario. $d$ is the distance between transmitter and receiver, $n$ is the distance dependent path loss exponent, and $X_\sigma (d)$  is the Gaussian distributed shadowing factor with zero mean in dB, and a variance of $\sigma^2$ in dB. 
\begin{figure}[t]
\centering
\includegraphics[width=1 \linewidth]{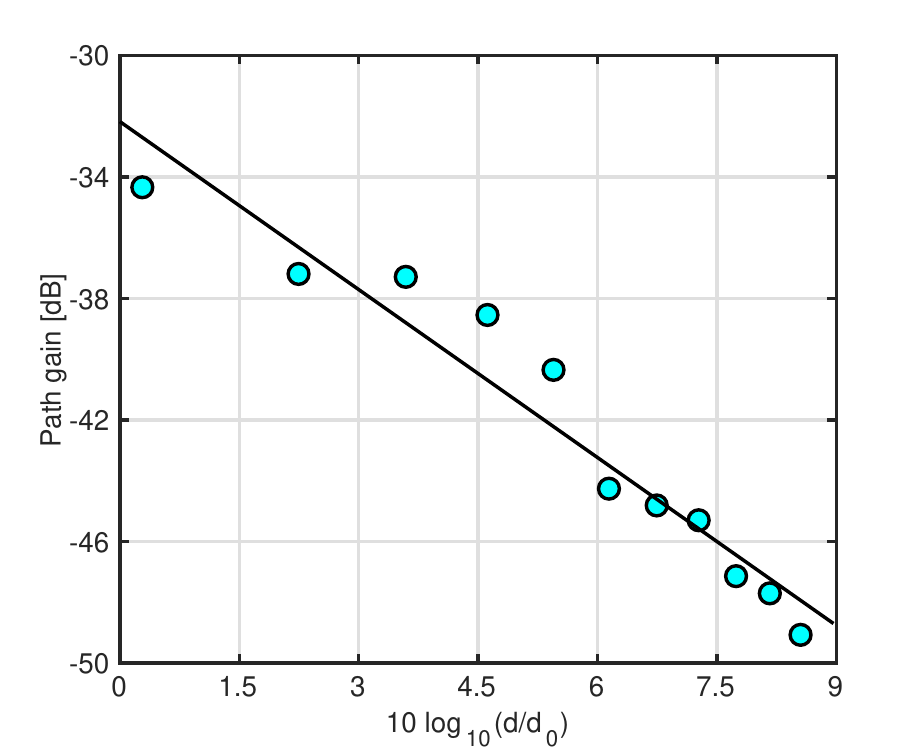}
 \caption{Linear fitting to determine distance dependent path loss. }
\label{PowStat}
\end{figure}
The path gain as a function of distance is plotted in Fig. \ref{PowStat}. A linear fitting model is used to find the slope of the plot which is nothing but the path loss factor $n$. The path loss factor in our anechoic scenario\footnote{The path loss factor derived in this work is specific to the measured data in this work and will require extensive set of measurements to be generalized for any environment as such.} using this linear fitting model was found to be 1.843. Similar close values have been reported in different indoor environments\cite{indoorPathloss,indoorPathloss2,pathLoss3}. This path loss modeling shows the applicability of such novel platform for mmWave channel characterizations.

\section{Conclusion} \label{conclusion}
In this work, a novel mmWave channel emulation method to analyse the Doppler spread, and path loss modeling are presented and discussed in detail. The UAV motion was emulated by a robotic arm considering the effect of wind gusts, which was modelled by the famous Dryden wind model. Doppler spread in both positive and negative frequency axis were determined at the noise floor of -60 dB in an anechoic chamber.  A total of 11 distance points were considered, and  in average, the Doppler spread was found to be around +20 and -20 Hz. No-motion data points were then selected to determine the path loss exponent for this anechoic chamber environment, which in our scenario came out to be 1.843.  These results and the application of such novel platform to study next generation wireless communication systems will definitely help propel the development, and design of such UAV assisted communication systems in the near future. As a future work, multi-beam phase array antenna will be considered for 3D scattering channel models with this type of emulation setup. Also, depolarization effect because of wind, attenuation and scattering over time is in the future scope of this work.


%



\section*{Acknowledgment}

The authors would like to thank Dr. Charles Bunting and Dr. Mehdi Bahadorzadeh for their great support in conducting measurements in the anechoic chamber. Also, the authors would like to thank Mohammed Musaqlab, Paul Hodgden,  and Scott Walker for their kind and generous help during the measurement campaign.  This work was supported in part by the Qatar National Research Fund under Grant NPRP13S-0130-200200 (a member of The Qatar Foundation), and  in part by the National Science Foundation (NSF), USA under award number 1925147.

\ifCLASSOPTIONcaptionsoff
  \newpage
\fi



%
\bibliographystyle{IEEEtran}

\bibliography{biblo}

%


\begin{IEEEbiography}[{\includegraphics[width=1in,height=1.25in,keepaspectratio]{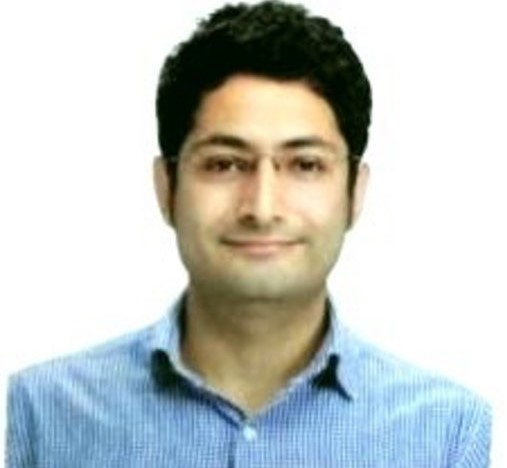}}]{Amit Kachroo (S'14)}  received his B.Tech degree in Electronics and Communications Engineering from NIT, Srinagar, India (2005-2009), and M.Sc degree from Istanbul Sehir University, Istanbul, Turkey (2015-2017) respectively. From 2010 to 2014, he worked as a project engineer with Nokia Networks, India on GSM, WCDMA, LTE, and RF technologies. In 2017, he joined Oklahoma State University, Stillwater, USA  to pursue  his Ph.D. degree in Electrical and Computer Engineering. In summer of 2020, he worked  as an Applied Scientist Intern at Wireless Technology Group  of Amazon Lab126, California, USA in the domain of  IoT  localization with AI/ML technology. Prior to that, in summer of 2019, he worked  as a software intern at Tensilica R\&D group of Cadence Design Systems, California, USA in  mmWave radar technology.  His research interest are in statistical modeling of wireless channels, mmWave channel modeling, statistical learning, machine learning, and cognitive radios.  He  also serves as a reviewer in  IEEE Communications Magazine, IEEE Transactions on Wireless Communications, IEEE JSAC, IEEE Access, and IEEE VTC Conference among others.  
\end{IEEEbiography}

\begin{IEEEbiography}[{\includegraphics[width=1in,height=1.25in,keepaspectratio]{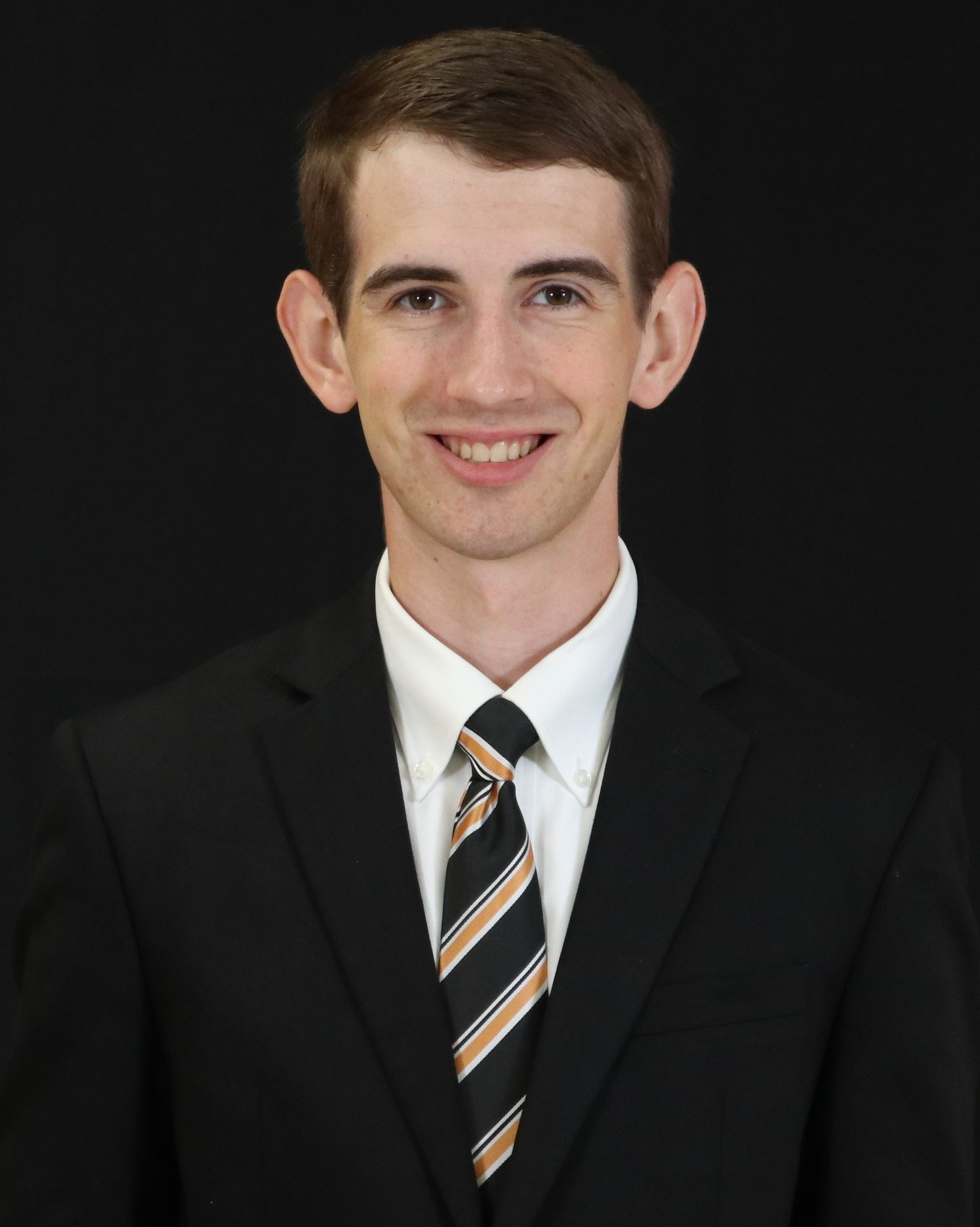}}]{Collin A. Thornton}   is a Honors College junior in the Department of  Electrical and Computer Engineering at Oklahoma State University. He joined OSU as both a CEAT Scholar and a Freshman Research Scholar, and he is pursuing dual degrees in Electrical and Computer Engineering. He was selected as an Institutional Nominee for the Goldwater Scholarship during his sophomore year. During his first semester at OSU, he began work in the Control, Robotics, and Automation lab in the Department of Mechanical and Aerospace Engineering. His work focuses on the development and application of path planning algorithms in unmanned aerial vehicles.  
\end{IEEEbiography}

\begin{IEEEbiography}[{\includegraphics[width=1in,height=1.25in,keepaspectratio]{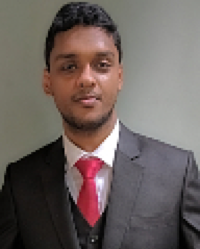}}]{Md Arifur Sarker}  received the B. Sc degree in Electrical \& Electronic Engineering from Rajshahi University of Engineering \& Technology, Rajshahi, Bangladesh in 2017. He is currently working toward the Ph. D. degree in Electrical and Computer Engineering at Oklahoma State University, Stillwater, OK, USA. He is currently with Systems on New Integrated Circuits (SONIC) group. His current research interests include radiation effects in RF circuits, phase array antenna and radar system and wideband circuit design for future communication system.  
\end{IEEEbiography}

\begin{IEEEbiography}[{\includegraphics[width=1in,height=1.25in,keepaspectratio]{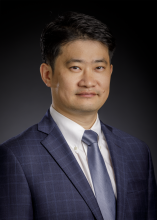}}]{Wooyeol Choi (M’11-SM’19)} received the B. S. degree in Electronic Engineering from Yonsei University, Seoul, Korea, the M. S. and the Ph. D. degrees in Electrical Engineering from Seoul National University, Seoul, Korea. From 2011 to 2018, he was with the Texas Analog Center of Excellence (TxACE) at the University of Texas at Dallas, Richardson, Texas, USA, first as a Research Associate later as an Assistant Research Professor. From 2018, he has been an Assistant Professor in the School of Electrical and Computer Engineering at Oklahoma State University, Stillwater, Oklahoma, USA. His research is focused on design and characterization of integrated circuits and systems for various applications from RF to terahertz frequency range.
\end{IEEEbiography}

\begin{IEEEbiography}[{\includegraphics[width=1in,height=1.25in,clip,keepaspectratio]{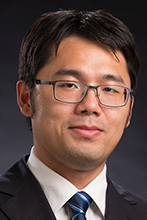}}]{He Bai (M'15)} received his B.Eng. degree from the Department of Automation at the University of Science and Technology of China, Hefei, China, in 2005, and the M.S. and Ph.D. degrees in Electrical Engineering from Rensselaer Polytechnic Institute in 2007 and 2009, respectively. From 2009 to 2010, he was a Post-doctoral Researcher at the Northwestern University, Evanston, IL. From 2010 to 2015, he was a Senior Research and Development Scientist at Utopia Compression Corporation. He was the Principal Investigator for a number of research projects on sense-and-avoid, cooperative target tracking, and target hand off in GPS-denied environments. In 2015, he joined the Mechanical and Aerospace Engineering Department at Oklahoma State University as an assistant professor. He has published over 70 peer-reviewed journal and conference papers related to control and robotics and a research monograph ``Cooperative control design: a systematic passivity-based approach" in Springer. He holds one patent on monocular passive ranging. His research interests include multi-agent systems, nonlinear estimation and sensor fusion, path planning, intelligent control, and GPS-denied navigation.
\end{IEEEbiography}

\begin{IEEEbiography}[{\includegraphics[width=1in,height=1.25in,clip,keepaspectratio]{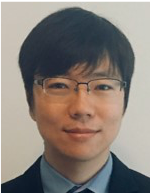}}]{Ickhyun Song (S’07-M’11) } received his B.S. and M.S. degrees in electrical engineering from Seoul National University, Seoul, Republic of Korea, in 2006 and 2008, respectively, and his Ph.D. degree in electrical and computer engineering from Georgia Institute of Technology, Atlanta, GA in 2016. 
From 2008 to 2012, he was Design Engineer at Samsung Electronics, Hwasung, Republic of Korea, where he contributed to the development of next-generation memory products. From 2017 to 2018, he was Research Engineer at Georgia Institute of Technology. In 2018, he joined the faculty at Oklahoma State University, and is currently Assistant Professor in the School of Electrical and Computer Engineering. His research interest includes extreme-environment electronics, radiation effects, RF/millimeter-wave devices and circuits, and CMOS/SiGe HBT device physics.	Dr. Song was a recipient of the 2007-2008 Samsung Semiconductor Scholarship, the Silver Paper Award at the 2007 IEEE Seoul Section Student Paper Contest, the Gold Prize at the 2008 Samsung HumanTech Paper Award, the 2012-2013 Fulbright Graduate Study Award, and the 2016 GEDC (Georgia Electronic Design Center) Best Poster Award.
\end{IEEEbiography}

\begin{IEEEbiography}[{\includegraphics[width=1in,height=1.25in,clip,keepaspectratio]{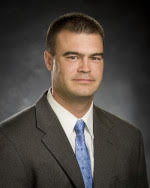}}]{John F. O'Hara (SM'19)} received his BSEE degree from the University of Michigan in 1998 and his Ph.D. (electrical engineering) from Oklahoma State University in 2003.  He was a Director of Central Intelligence Postdoctoral Fellow at Los Alamos National Laboratory (LANL) until 2006.  From 2006-2011 he was with the Center for Integrated Nanotechnologies (LANL) and worked on numerous metamaterial projects involving dynamic control over chirality, resonance frequency, polarization, and modulation of terahertz waves.  In 2011, he founded a consulting/research company, Wavetech, LLC.  In 2017, he joined Oklahoma State University as an Assistant Professor, where he now studies metamaterials, terahertz communications, photonic sensing methods, and Internet of Things applications.   He has around 100 publications in journals and conference proceedings.
\end{IEEEbiography}

\begin{IEEEbiography}[{\includegraphics[width=1in,height=1.25in,clip]{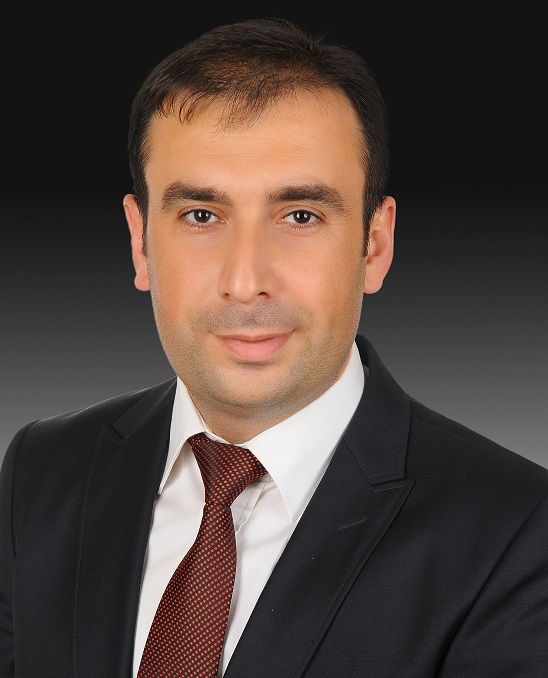}}]{Sabit Ekin (M'12)}  received the B.Sc. degree in electrical and electronics engineering from Eski\c sehir Osmangazi University, Turkey, in 2006, the M.Sc. degree in electrical engineering from New Mexico Tech, Socorro, NM, USA, in 2008, and the Ph.D. degree in electrical and computer engineering from Texas A\&M University, College Station, TX, USA, in 2012. He was a Visiting Research Assistant with the Electrical and Computer Engineering Program, Texas A\&M University at Qatar from 2008 to 2009. In summer 2012, he was with the Femtocell Interference Management Team in the Corporate Research and Development, New Jersey Research Center, Qualcomm Inc. He joined the School of Electrical and Computer Engineering, Oklahoma State University, Stillwater, OK, USA, as an Assistant Professor, in 2016. He has four years of industrial experience from Qualcomm Inc., as a Senior Modem Systems Engineer with the Department of Qualcomm Mobile Computing. At Qualcomm Inc., he has received numerous Qualstar awards for his achievements/contributions on cellular modem receiver design. His research interests include the design and performance analysis of wireless communications systems in both theoretical and practical point of views, interference modeling, management and optimization in 5G, mmWave, HetNets, cognitive radio systems and applications, satellite communications, visible light sensing, communications and applications, RF channel modeling, non-contact health monitoring, and Internet of Things applications.
\end{IEEEbiography}




\end{document}